\documentstyle[preprint,aps,eqsecnum]{revtex}
\tightenlines
\def\m@thcombine#1#2{%
  \setbox0=\hbox{$#1$}
  \setbox1=\hbox{$#2$} 
  \ifdim\wd0>\wd1
    \setbox0=\hbox to\wd1{\hss\box0\hss}
  \else
    \setbox1=\hbox to\wd0{\hss\box1\hss}
  \fi
  \mathop{\vcenter{
    \offinterlineskip\box0\box1}}}
\def\lesim{\m@thcombine<\sim}
\def\gesim{\m@thcombine>\sim}

\begin{document}

\draft
\title{ YANG-MILLS VACUUM INSTABILITY DUE TO INFRARED FINITE LATTICE GLUON PROPAGATOR }

\author{ M. Priszny\'ak }

\address{HAS, CRIP, RMKI, Theory Division, P.O.B. 49, H-1525 Budapest 114, Hungary \\ email address: prisz@sunserv.kfki.hu}

\maketitle

\begin{abstract} 
Using the effective potential approach for composite operators, we have analytically
evaluated the truly nonperturbative vacuum energy density as obtained by using a model infrared
finite gluon propagator which was suggested by lattice simulations. 
The truly nonperturbative vacuum
 energy density is defined as the truly nonperturbative part of the full gluon
propagator integrated over the deep infrared region (soft momentum region). 
With this defition, this is a manifestly
gauge invariant quantity. We have explicitly shown that the corresponding effective potential has 
always an imaginary part which means that the vacuum of this model is unstable.
Thus this model cannot have a true ground state.  
\end{abstract}


\vfill

\eject

\section{Introduction}

In recent papers [1,2]  a general method (prescription) was formulated how to
calculate correctly the truly nonperturbative Yang-Mills (YM) vacuum energy density (VED) of the ground state in QCD
models by using the effective potential 
approach for composite operators [3,4]. This allows one to investigate the vacuum
structure by substituting some well justified ansatz for the full gluon propagator 
\footnote{Recent brief reviews on both the continuum and lattice gluon propagators can be found in Refs. [5,6].}
since in the absence of external sources the corresponding effective 
potential is nothing else but the VED itself. The truly nonperturbative 
VED was defined as 
the truly nonperturbative part of the full
gluon propagator integrated over the deep infrared (IR) region (soft momentum  
region), i.e., the method assumes
that all kinds of perturbative contributions must be subtracted from the VED   
which, in general, is badly divergent. 
In order to factorize the scale dependence, the effective potential was introduced at a fixed scale.
This makes it possible to investigate the structure of the YM vacuum in terms  
of dimensionless
but physically meaningful variables and parameters. The nontrivial
minimization procedure, which can be performed in two different ways,\footnote{Both lead to the same numerical value of the truly nonperturbative    
VED provided it exists.} allows one to  determine the value of the soft cutoff 
as a function
of the corresponding scale parameter. This latter is inevitably present in any 
nonperturbative
model (either classical, quantum or lattice) of the full gluon propagator.    
If the chosen ansatz for the full gluon propagator is realistic then the
method will determine uniquely the truly nonperturbative VED, which should be  
always      
finite, negative and it should have no imaginary part (ee Ref. [1,7,8]).  Let  
us emphasize that the truly nonperturbatibe VED,
provided it is correctly calculated,  should be  either negative (stable vacuum
configurations if it also has nontrivial minimum or minima)
or  complex (which definetely signals unstable vacuum). 

Why is it so important to calculate the truly nonperturbative VED from first principles? 
First of all, this quantity is important in its own right since it is, by definition, nothing but the
bag constant (ie. the bag pressure) apart from the sign [9].
 Through the trace anomaly relation [10] it
assists in the correct estimation of such an important
phenomenological nonperturbative parameter like the gluon condensate as introduced in 
the QCD sum rules approach to resonance physics [11]. Furthermore, it helps 
in the resolution of the $U(1)$ problem [12] via the Witten-Veneziano (WV) formula for the 
mass of the $\eta'$ meson [13]. 
The problem here is that the topological susceptibility needed for this
purpose [14,15] is determined by the two point correlation function from 
which the perturbative contributions
should be correctly subtracted [13-18]. The same holds true for the above-mentioned bag constant which is 
a much more general quantity than the string tension since it is relevant for  
light quarks as well. Thus to correctly calculate 
the truly nonperturbative VED means to correctly understand the
structure of the QCD vacuum in different models.                               
           
In our previous publication [19] (see also the references therein) we have     
already investigated the structure of 
the classical vacuum in the Abelian Higgs model of the dual QCD ground state.  
We have explicitly shown that the vacuum of this model without string and  
with string contributions is unstable against quantum corrections. 

Now the main purpose of this work is to investigate the YM vacuum structure by 
using an IR finite (IRF) gluon propagator as suggested by lattice simulations  
in Ref. [20]. Still, let us display a few necessary definitions, to be used    
later, beforehand.

\section{The truly nonperturbative VED }

The relevant expression for the truly nonperturbative YM VED as derived (prescribed) in Ref. [1] is (in Euclidean metrics)

\begin{equation}
\epsilon^{np}_g = {1 \over \pi^2} \int_0^{q^2_0} dq^2 \ q^2 \left[ {3 \over 4}d^{NP}(q^2) - \ln [1 + 3 d^{NP}(q^2)]\right],
\end{equation}
where $q_0^2$ is the above-mentioned soft cutoff which separates the deep IR region 
(where the nonperturbative dynamics becomes dominant) from the perturbative region (see below). 
The truly nonperturbative gluon form factor is defined as follows:

\begin{equation}
d^{NP}(q^2, \Lambda_{NP}) = d(q^2, \Lambda_{NP}) - d(q^2, \Lambda_{NP} =0).
\end{equation}
Here $\Lambda_{NP}$ is the also above-mentioned scale parameter that is responsible for
the nonperturbative dynamics in the model under consideration. This definition 
explains the difference between the truly nonperturbative part $d^{NP}(q^2)$  
and the full gluon propagator $d(q^2)$ which is nonperturbative itself.        
Moreover, it guarantees that
the truly nonperturbative VED (2.1) is a manifestly gauge  
invariant quantity. Though the full gluon propagator is explicitly gauge       
dependent, its truly nonperturbative part is not like that since the explicit  
gauge dependence (in fact the longitudinal term which is not renormalized) vanishes along with the perturbative terms contained in $d(q^2, \Lambda_{NP} =0) \equiv d^{PT}(q^2)$.
(The longitudinal part, which explicitely depends on the gauge, is always perturbative.)  
It is easy to see that by "PT" we mean the intermediate (IM) and the         
ultraviolet (UV) regions (the IM region remains $terra \  
incognita$ in QCD). Fortunately the "PT" part is of no importance here.        
                                                      
Thus the separation of "NP vs. PT" becomes exact because of the definition     
(2.2). The separation of "soft vs. hard" momenta also becomes exact because of 
the minimization procedure. The analysis of the truly nonperturbative 
VED (2.1) after the above-mentioned scale factorization provides, in addition, 
an exact criteria to distinguish  between stable and unstable vacuum.     
Thus the truly nonperturbative VED, as it is given in Eq. (2.1), is uniquely
defined. It is truly nonperturbative since it contains no perturbative information at all.
This is rather similar to the lattice approach where, by using different 
"smoothing" techniques such as "cooling" [21], "cycling" [22], etc., it is possible 
to "wash out" all type of  perturbative fluctuations and excitations of
the gluon field configurations from the QCD vacuum in order to deal only with  
the true nonperturbative structure.                                            

The above briefly-described general method [1,2] can serve as a test of        
different quantum, classical as well as lattice models of QCD.

\section{MMS IRF gluon propagator}

Let us investigate the quantum structure of YM vacuum with the 
IRF behaviour of the full gluon propagator (in the Landau gauge) as suggested  
by lattice calculations in
Ref. [20] by Marenzoni, Martinelli and Stella (MMS). The propagator  
was parametrized as follows (from the very beginning in our notations and      
Euclidean metrics):

\begin{equation}
d(q^2) = { q^2 \over M^2 + Z q^2 (q^2 a^2)^{\eta} }.
\end{equation}
Here $M$ is the mass scale parameter which is responsible for the nonperturbative dynamics in this
model, i. e., $M= \Lambda_{NP}$ in our notation.
When the parameter $M$ formally goes to zero, only the perturbative part       
remains. The best estimates for the parameters $M$ and $a$ are $M = 160 \ MeV$ and $a^{-1} \approx 2.0 \ GeV$ (the inverse lattice spacing). The exponent is  
$\eta \approx 0.53$ (this mimics an anomalous dimension for the gauge field)   
and the renormalization constant is $Z \approx 0.1$.
Let us remind the reader that within the general method [1,2], the nonperturbative scale  
parameter is considered free, i.e., as "running" (when it formally goes
to zero then only the perturbative phase survives in the model) and its        
numerical value will
be used only, provided it exists at all,  at the final stage in order to       
numerically evaluate the corresponding truly nonperturbative VED.   
The subtraction procedure in (2.2) now looks like 

\begin{equation}
d^{NP}(q^2) = d(q^2, M^2) - d(q^2, M^2=0) = - { M^2 \over Z(q^2 a^2)^{\eta} [M^2 + Z q^2 (q^2 a^2)^{\eta}] }.
\end{equation}

\subsection{Fixing the lattice spacing}

In order to factorize the scale dependence in (2.1), let us first choose the   
lattice spacing $a$ as an extra scale parameter [1]. Then $d^{NP}(q^2)$ in     
Eq. (3.2) becomes

\begin{equation}
d^{NP}(z, b) = - { 1  \over Z z^{\eta} [1+ Z b^{-1} z^{1 + \eta}] },
\end{equation}
where 
                           
\begin{equation}
z = q^2 a^2, \quad z_0 = q_0^2 a^2, \quad b= M^2 a^2.                        
\end{equation}
Here $z_0$ is the corresponding dimensionless soft cutoff  while the
parameter $b$ has a very clear physical meaning, i. e., when it is zero (which 
means that $M^2 \rightarrow 0$) only the perturbative phase  remains.

Substituting (3.3) and (3.4) into the effective potential (2.1), one obtains

\begin{equation}
\Omega_g (z_0, b) = a^4 \epsilon_g^{np} = - {1 \over \pi^2} \Bigl\{ I_1(z_0, b) + I_2(z_0, b) \Bigr\},
\end{equation}
where the integrals are given as follows:

\begin{eqnarray}
I_1 (z_0, b) &=& \int \limits_0^{z_0} dz\, z\, \ln \Big( 1 - { 3 \over Z z^{\eta}[1 + Z b^{-1} z^{1+\eta} ] } \Big),  \nonumber\\
I_2 (z_0, b) &=& { 3 \over 4 Z} \int \limits_0^{z_0} dz\, {z^{1-\eta} \over 1 +Z b^{-1} z^{1+\eta}} =  { 3 z_0^{2-\eta} \over 4 Z (2 - \eta)}
{}_2F_1 \Big( 1, {2 - \eta \over 1 + \eta}; 
{2 - \eta \over 1 + \eta} +1; - Z {z_0^{1 + \eta} \over b} \Big), 
\end{eqnarray}
and ${}_2F_1 ( 1, ... ; ... ; - Z z_0^{1 + \eta} /b)$ denotes the hypergeometrical function.

From these expressions it is possible to show that the effective potential (3.5) 
in the perturbative limit ($b \rightarrow 0$) vanishes indeed, while at infinity ($b \rightarrow \infty$)
 to-leading order becomes constant and it depends on $Z$
, $z_0$ and $\eta$. So at first sight, it seems the effective potential at a   
fixed scale (3.5) as function of the parameter $b$ may have a nontrivial       
minimum.                                                                       
   
Note however that  the first integral in Eqs. (3.6) 
always exhibits an imaginary part at any finite values of $z_0$ and $b$.      
To show this it suffices to investigate the function under logarithm in        
Eqs. (3.6)   

\begin{equation}
R \equiv R(z, Z, b, \eta) = 1 - { 3 \over Z z^{\eta}[ 1 + Z b^{-1} z^{1 +\eta}]}.
\end{equation}
 Let us notice that $R(z=0)= - \infty$ holds true for $any \ fixed$ $Z$,
$b$ and $\eta$. Since $R$ is regular as a function of $z$ in the whole open interval $(0, z_0]$ for 
any $Z$, $b$ and $\eta$\footnote{The poles at $1 + Z b^{-1}
z^{1 +\eta} = 0$ do not lie in the interval $[0, z_0]$ since $z_0$ is always 
positive, by definition.}, it simply follows from the Boltzano-Weierstrass     
theorem that there exists an open interval (with $z=0$ as the left open end point)
 where $R$ is negative, provided 
$R$ becomes non-negative somewhere in the interval $(0, z_0]$. If this were not
true then $R$ must be negative in the whole interval  $(0, z_0]$. Having such  
an interval where $R<0$, and taking
into consideration that the logarithm is a monotonous function,
 we certainly have an imaginary part in the effective potential (3.5) for any  
finite set of parameters $Z, \ b, \ \eta, \ z_0$.  

Let us derive also the formal  "stationary" condition with
respect to $b$, namely $\partial \Omega_g (z_0, b) / \partial b =0$. Then from 
Eqs. (3.5) and (3.6), one obtains                                      
             
\begin{equation}
\int \limits_0^{z_0} dz\, z^2\, {Z \varphi(z,b) z^{\eta} + 1 \over \varphi^2(z,b) [Z \varphi (z,b)z^{\eta} -3]} = 0
\end{equation}
with

\begin{equation}
\varphi(z,b) = 1 +Z b^{-1} z^{1+\eta}. 
\end{equation}
From the  estimates for the parameters $M$ and $a$ mentioned above, 
it follows that $b \ll 1$. Still we use $b \leq 1$ in order to be sure that we do not miss the
nontrivial minima. Numerically evaluating the "stationary" condition (3.8)
(on account of the numerical values $Z \approx 0.1$ and $\eta \approx 0.53$), nevertheless
 we  found 
that  only the trivial solution $z_0=0$ exists indeed.        

Thus the vacuum of this model is unstable and therefore it is physically not acceptable. 
The MMS IRF gluon model propagator can be related neither to quark
confinement nor to dynamical chiral symmetry breakdown (DCSB). 
This statement is in complete agreement with the conclusion given in Ref. [23].
There it was  
shown that the MMS gluon propagator neither confines quarks nor it breaks   
chiral symmetry dynamically. Three special type of expressions for the dressed 
quark-gluon vertex 
(free from ghost contributions) were used in their investigation of the quark  
SD equation. Our result is, however, more general since   
we do not require a particular choice for the dressed quark-gluon vertex.

\subsection{Fixing the soft cutoff }

It is instructive to further investigate the presently discussed model by      
choosing the dimensionless variables and parameters as follows: 

\begin{equation}
z={q^2 \over M^2}, \quad  z_0={q_0^2 \over M^2}.    
\end{equation}
For simplicity's sake, we use the same notations for the
dimensionless set of variables and parameters as in Eq. (3.4). Now when
the parameter $z_0$ goes to infinity (at a fixed  soft cutoff, see below)   
then only the perturbative phase survives  
($M \rightarrow 0$). In this case, $d^{NP}(q^2)$ in Eq. (3.2) becomes

\begin{equation}
d^{NP}(z) = - { 1  \over a_1 z^{\eta} [1+ a_1 z^{1 + \eta}] },
\end{equation}
where 

\begin{equation}
a_1 = Z (Ma)^{2\eta}.
\end{equation}                                                                 
 Substituting this into the Eq. (2.1) and fixing the soft cutoff itself [1],   
one obtains

\begin{equation}
\bar \Omega_g (z_0; a_1) = { 1 \over  q_0^4} \epsilon_g^{np} = - {1 \over \pi^2} \times z_0^{-2} \Bigl\{ I_1(z_0; a_1) + I_2(z_0; a_1) \Bigr\},
\end{equation}
where the integrals are given as follows:

\begin{eqnarray}
I_1 (z_0; a_1) &=& \int \limits_0^{z_0} dz\, z\, \ln \Big( 1 - { 3 \over a_1 z^{\eta}[1 + a_1 z^{1+\eta} ] } \Big),  \nonumber\\
I_2 (z_0; a_1) &=& { 3 \over 4 a_1} \int \limits_0^{z_0} dz\, {z^{1-\eta} \over 1 + a_1 z^{1+\eta} }. 
\end{eqnarray}
A specific feature of this model is that the combination 

\begin{equation}
a_1 z_0^{\eta} = Z (Ma)^{2\eta} (q_0^2 M^{-2})^{\eta} = Z (q_0^2 a^2)^{\eta} = 
\nu
\end{equation}
is fixed when the soft momentum cutoff $q_0$ is fixed (like in this case, see Eq. (3.12)). Thus the
effective potential (3.13) and corresponding integrals
(3.14) become more complicated functions of $z_0$, namely

\begin{equation}
\bar \Omega_g (z_0, \nu) = { 1 \over  q_0^4} \epsilon_g^{np} = - {1 \over \pi^2} \times z_0^{-2} \Bigl\{ I_1(z_0, \nu) + I_2(z_0, \nu) \Bigr\},
\end{equation}
where integrals are given now as follows:

\begin{eqnarray}
I_1 (z_0, \nu) &=& \int \limits_0^{z_0} dz\, z\, \ln \Big( 1 - { 3 \over \nu z_0^{-\eta} z^{\eta}[1 + \nu z_0^{-\eta} z^{1+\eta} ] } \Big),  \nonumber\\
I_2 (z_0, \nu) &=& { 3 z_0^{\eta} \over 4 \nu} \int \limits_0^{z_0} dz\, {z^{1-\eta} \over 1 + \nu z_0^{-\eta} z^{1+\eta} } = 
{ 3 z_0^2 \over 4 \nu (2 - \eta)} {}_2F_1 \Big( 1, {2 - \eta \over 1 + \eta}; 
{2 - \eta \over 1 + \eta} +1; - \nu z_0 \Big),
\end{eqnarray}
and ${}_2F_1 ( 1, ... ; ... ; - \nu z_0)$ is the hypergeometrical
function. 

The dependence on the parameter $z_0$ becomes  more complicated than in        
Eq. (3.13) indeed. 
Nevertheless, like in the previous case, it is possible to show again,
by analysing the first integral in Eqs. (3.17), 
 that the effective potential
(3.16) will have an imaginary part (at any finite values  
of parameters $z_0$ and $\nu$). 
Consequently the vacuum arising from this lattice model gluon propagator is    
unstable indeed.

\acknowledgements

The author would like to thank V. Gogohia for suggesting this investigation as 
well as for many useful discussions and remarks.

\vfill

\eject

\end{document}